\documentclass[aps,prl,twocolumn,superscriptaddress,groupedaddress,nofootinbib]{revtex4}
\pdfoutput=1
\usepackage{graphicx}
\usepackage{dcolumn}   
\usepackage{bm}        

\usepackage{amsmath} 
\usepackage{graphicx}
   \usepackage{multirow}

\usepackage{epstopdf}
\newcommand{\PRE}[1]{{#1}} 

\newcommand{\be}{\begin{equation}}
\newcommand{\ee}{\end{equation}}
\newcommand{\bea}{\begin{eqnarray}}
\newcommand{\eea}{\end{eqnarray}}

\newcommand{\nn}{\nonumber}

\def\gev{\, {\rm GeV}}

\newcommand{\gsim}{\lower.7ex\hbox{$\;\stackrel{\textstyle>}{\sim}\;$}}
\newcommand{\lsim}{\lower.7ex\hbox{$\;\stackrel{\textstyle<}{\sim}\;$}}

\newcommand{\cm}{{\rm cm}}

\newcommand{\fermi}{{\emph{Fermi}}-LAT}

\newcommand{\braket}[1]{\left<#1\right>}

\usepackage{enumerate}
\usepackage{amssymb,amsmath}

\begin{document}

%


\title{
\PRE{\vspace*{1.3in}}
\textsc{Searching for light from a dark matter clump}
\PRE{\vspace*{0.1in}}
}

\author{Tathagata Ghosh}
\affiliation{\mbox{Department of Physics and Astronomy, University of
Hawai'i, Honolulu, HI 96822, USA }
}

\author{Jason Kumar}
\affiliation{\mbox{Department of Physics and Astronomy, University of
Hawai'i, Honolulu, HI 96822, USA }
}

\author{Danny Marfatia}
\affiliation{\mbox{Department of Physics and Astronomy, University of
Hawai'i, Honolulu, HI 96822, USA }
}

\author{Pearl Sandick}
\affiliation{\mbox{Department of Physics and Astronomy, University of Utah, Salt Lake City, UT  84112, USA}
\PRE{\vspace*{.1in}}
}

\begin{abstract}

The DAMPE experiment has recently reported an electron spectrum that can be explained
by dark matter annihilation into charged lepton pairs in a nearby
dark matter clump. The accompanying bremsstrahlung may yield a gamma-ray excess with a known spectral shape that
extends over an angular scale of ${\cal O}(10^\circ)$.  
We show that such an excess is not present in \fermi~data.
\end{abstract}

\maketitle


\section{Introduction}

The DArk Matter Particle Explorer (DAMPE) has recently acquired new data on the cosmic ray
electron plus positron spectrum, indicating a spectral break at around $0.9$~TeV
and a possible peak at $\sim 1.5$~TeV~\cite{Ambrosi:2017wek}.  A potential explanation for the peak is the annihilation
of $\sim 1.5$~TeV dark matter (DM) particles to electron-positron pairs ($XX \rightarrow e^+ e^-$) in a nearby dark matter clump~\cite{Yuan:2017ysv}.
If this
explanation is correct, then an associated flux of photons from final state radiation (FSR) ($XX \rightarrow e^+ e^- \gamma,
e^+ e^- Z, e^+ W^- \nu_e$, etc.) is expected.
The resulting gamma-ray spectrum is well defined
and points back to the source.  
We perform a search of the Fermi Large Area Telescope (\fermi) data 
for evidence of such a dark matter clump.

The excess at $1.5$~TeV reported by DAMPE consists of events in a single energy bin.  For this excess to
be produced by DM annihilation, the mass of the DM particle must be close
to $1.5$~TeV with a substantial branching ratio (BR) to the $e^+ e^-$ final state.  Furthermore, the
source must be within about 0.3~kpc of the solar system so that the electron spectrum is not softened by cooling processes such as synchrotron emission~\cite{Yuan:2017ysv}.
Even for a nearby source,
there must be an enhancement of the annihilation rate beyond that expected from a thermal DM 
particle annihilating in a smooth DM halo.  Possibilities include the Sommerfeld enhancement of DM annihilation, a local DM
overdensity encompassing the solar system, or a nearby DM clump.  We focus on the case of a nearby
clump, in which case the photon emission from DM annihilation is expected to be highly directional, and 
therefore distinguishable from the isotropic background.  A benchmark example of a dark matter clump capable of
producing the DAMPE excess is a clump of core size 10 pc, with an overdensity of a factor of $\sim 1000$, located
about 100~pc away~\cite{Yuan:2017ysv}.  Such a clump needs to have
${\cal L}\equiv  \int dV \, \rho^2 \sim 3.5 \times 10^{64} \gev^2 / \cm^{3}$~\cite{Yuan:2017ysv}.

Gamma-ray constraints on the DM clump interpretation of the DAMPE excess were considered in Ref.~\cite{Yuan:2017ysv}.
In particular, photon emission arising from DM annihilation in the clump was integrated over a patch of radius $1^\circ$ on the sky,
and the resulting differential flux was compared to the 10-year \fermi~point source sensitivity.
It was found that the photon flux produced
from DM annihilation is below the \fermi~sensitivity by about an order of magnitude.  But a clump with a
dense core of size 10~pc located 100~pc away covers an angular size of ${\cal O}(10^\circ)$ on the sky, and the resulting factor ${\cal O}(100)$ increase
in flux could make such a clump detectable at \fermi.

Indeed, a clump at a distance of 100~pc with ${\cal L} = 3.5 \times 10^{64} \gev^2 / \cm^{3}$
has an average $J$-factor of $J_{clump} \approx {\cal O}(10^{24}) \gev^2 / \cm^{5}$, which is
about a factor of $100$ larger than the $J$-factor of
Draco, averaged over an angular size of $1^\circ$~\cite{Geringer-Sameth:2014yza}.
For $m_X \sim 1.5$~TeV, the bound on the DM annihilation cross section
to the $e^+ e^-$ channel obtained by \fermi~in a stacked dwarf analysis is about a factor $10^{2-3}$ larger than the
thermal cross section~\cite{Ackermann:2015zua}.  Of course, this is a stacked analysis of 15 dSphs, in which the flux from each is integrated over $1^\circ$.  But since the solid angle encompassing the DM clump could be as much as $100$~times larger than that for a dwarf,
a search for a photon excess from a DM clump in a $10^\circ$ region amounts to a stacked analysis of
${\cal O}(10^2)$ dwarf-sized objects, each of which has a $J$-factor ${\cal O}(100)$
times larger than that of a typical dwarf.  Thus, \fermi~may probe cross sections ${\cal O}(10^{2-3})$ smaller
than in the stacked dwarf analysis, thereby putting the thermal annihilation cross section within reach.  Moreover, if DM
also annihilates to other final states, then the photon signal could be even more striking.

Our search strategy for the putative dark matter clump is as follows. We cover the sky outside
the galactic plane with 144 regions of interest (ROIs) of equal solid angle.  In each ROI, we
fit the photon data to a background model including photon emission from identified point sources in the ROI, as well as the
isotropic background.  The residual is then  fit to the photon spectrum arising from the annihilation process, $XX \rightarrow \ell^+ \ell^-$
(including FSR). Finally, we compare the $\chi^2$ of the best fit scenario to the null hypothesis (no dark matter
annihilation).
We explore three annihilation  scenarios: \\
{\it Case I}: $X X    \rightarrow  e^+ e^-$ only;\\
{\it Case II}:  $X X   \rightarrow \ell^+ \ell^- \, (\ell = e, \mu)$, with each BR $=\frac{1}{2}$;\\
{\it  Case III}:  $X X   \rightarrow  \ell^+ \ell^- \, (\ell = e, \mu, \tau)$, with each BR $=\frac{1}{3}$.



\section{Data selection and background model}

We use 9.3 years of \fermi~ data from August 04, 2008 to December 02, 2017, which corresponds to \emph{Fermi} mission elapsed time, $239557418 - 533867602$~s. Since we are searching for extended sources, we select events with ``Pass 8 ultracleanveto" event class ({\tt evclass} = 1024), with {\tt evtype} = 3, using {\tt P8R2$\_$ULTRACLEANVETO$\_$V6} instrument response functions~\cite{class}. We analyze events in the energy range, $[0.5, \,500]$ GeV, with a maximum zenith angle of $90^\circ$, and consider the full sky except for the Galactic plane ($b=[-10^\circ, \, 10^\circ]$, where $b$ is the Galactic latitude).

We cover the sky with 144 ROIs, centered at Galactic coordinates $(b,l)$ given by $b= n*20^\circ$ ($n= \pm 1, ..., \pm 4$),
$\l = 10^\circ + m \times 20^\circ$ ($m = 0, ..., 17$).  Each ROI has a width of $20^\circ$ around the ROI center ($\Delta \Omega \approx 0.122$).
So all regions outside
the Galactic plane are sampled, but regions near the Galactic poles are oversampled.

For the purpose of modeling the background diffuse gamma rays, we employ templates released by the \fermi~collaboration with Pass 8 data~\cite{Acero:2016qlg}: 
{\tt gll$\_$iem$\_$v06.fits} for the Galactic interstellar emission model (IEM) and the corresponding {\tt iso$\_$P8R2$\_$ULTRACLEANVETO$\_$V6$\_$v06.txt} for the isotropic component. In addition, we employ the \fermi~ Third Source Catalog ({\tt 3FGL})~\cite{3FGL} to account for point source contributions to the background. 

To estimate the backgrounds for each ROI, we use the {\tt Fermipy$\_$v0.14.1} Python package~\cite{fermipy}, which in turn uses LAT {\tt ScienceTools$\_$v10r0p5}~\cite{LAT-ST}. Although each ROI is analyzed separately, we do include the {\tt 3FGL} point sources within a width of $30^\circ$ around each ROI center to account for spillover from point sources belonging to adjacent ROIs. We bin the data in 8 energy bins per decade with a $0.05^\circ$ pixel size for each ROI. The details of our background modeling are given in the Appendix. Once the background model for each ROI is optimized, we extract the recorded photon counts and expected model counts per energy bin for DM analyses.

\section{Photon signal from FSR and inverse Compton scattering}
For each annihilation channel, a prompt photon flux arises  from FSR and from the
decay of $\tau$ leptons in the final state (which yield neutral pions that in turn decay to photons).  The energy spectrum of this prompt
photon signal is
determined using the publicly available {\tt PPPC4DMID} code~\cite{PPPC, PPPC-2}. The FSR photon flux necessarily points directly back to the dark matter
clump.

The $e^\pm$ pairs produced by DM annihilation also undergo inverse Compton scattering (ICS) on the interstellar radiation field
(ISRF), potentially producing another source of photons arriving from the direction of the clump. To determine the photon flux arising from ICS,
we adopt a semi-analytical approach. First, we obtain the energy density distribution of the ISRF, including the cosmic microwave background (CMB), starlight, and the infrared background (IR), by using model M2 outlined in Table~2 of   Ref.~\cite{Delahaye:2010ji}. In model M2, the energy density distribution is calculated by extracting the spectral energy distribution (SED) of the ISRF components 
from {\tt GALPROP}~\cite{galprop}. We then average the SED  over a cylinder of 2~kpc radius and half-width centered on the Earth. The averaged SED is fit by gray-body spectra with energy densities $U_{\text{rad}_i}$ and temperatures $T_{0_i}$, where $i=\{ \text{CMB, IR, starlight}\}$.

Next, we calculate the halo function, $I(\lambda_D(E, E_S),\vec{x})$, for electrons by solving the diffusion equation by the method of images~\cite{Baltz:1998xv}. 
The halo function contains the information on $e^\pm$ propagation through the diffusion length $\lambda_D$; $E_S$ is the $e^\pm$ energy at the source. 
We use the diffusion coefficient and energy loss coefficient models of Ref.~\cite{Yuan:2017ysv}. The secondary electron spectrum after propagation is given by
\begin{equation}
\Psi_e (\vec{x},E) = \dfrac{\kappa}{\bar{b}(E)} \int_{E}^{\infty} dE_S \, I(\lambda_D(E, E_S),\vec{x}) \, \,\dfrac{dN_e}{dE}(E_S)  ,
\end{equation}
where $\bar{b}(E)$ is the energy loss coefficient, $\kappa =  \braket{\sigma v}/2m_X^2$ and $\braket{\sigma v}$ is the thermally averaged annihilation cross section.

Once the secondary electron spectrum is known, the ICS spectrum is determined by
\begin{eqnarray}
\dfrac{dN_{\text{ICS}}}{dE_{\gamma}} & = & \dfrac{2}{E_{\gamma}} \dfrac{1}{4 \pi} \int_{l.o.s} ds \, d\Omega \int_{E^{\text{min}}_e}^{m_X} dE_e \,  P_{\text{IC}}(E_{\gamma}, E_e) \nn \\
&& \times \, \Psi_e (\vec{x},E_e) \, .
\end{eqnarray}
We use the Klein-Nishina limit of the ICS emission spectra $P_{\text{IC}}$ (see Ref.~\cite{PPPC-2} for details).

\begin{figure}[t]
\centering
\includegraphics[scale=.20]
{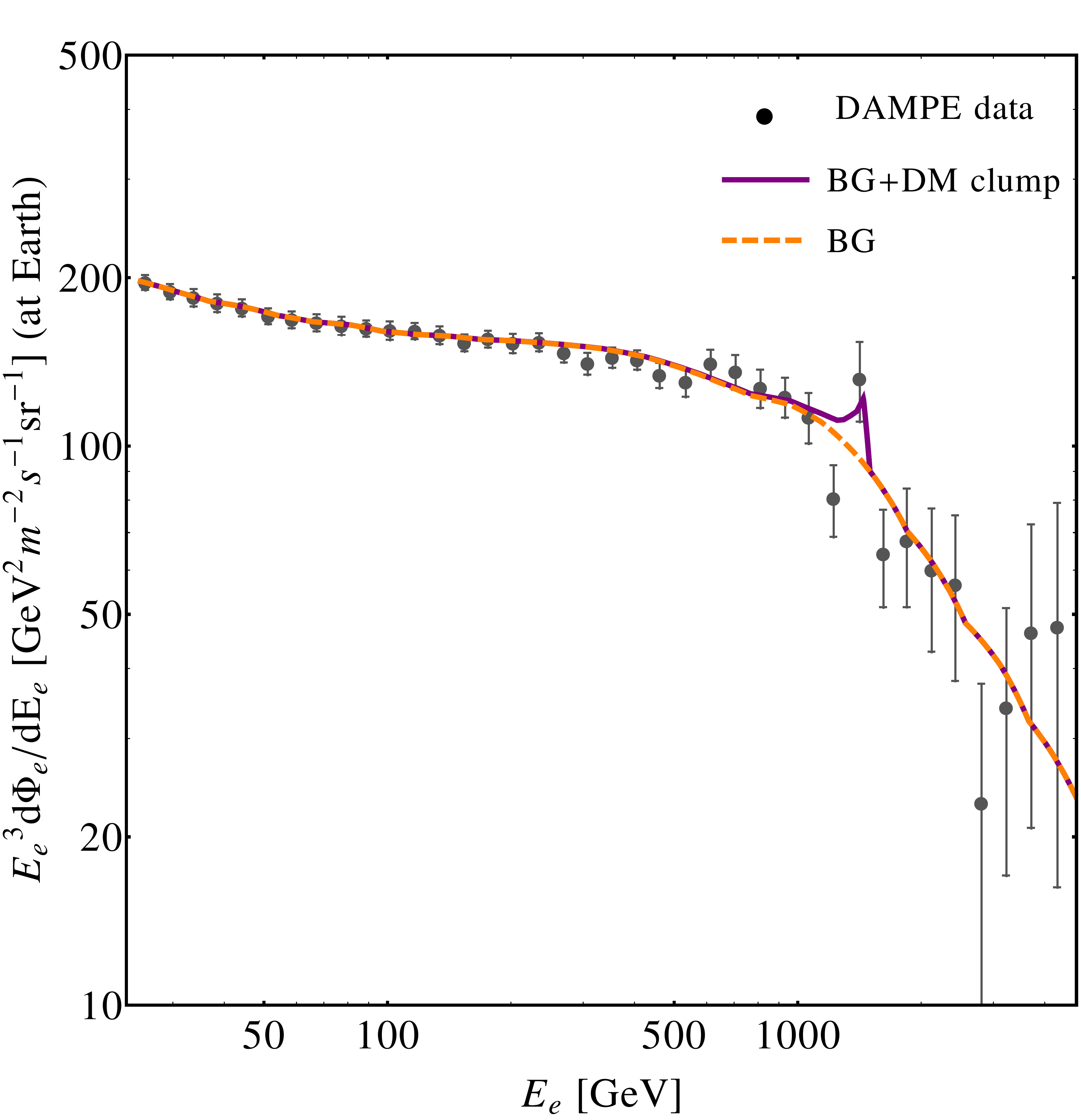}
\caption{Secondary $e^++e^-$ spectrum originating from a DM clump at a distance of 0.1~kpc and centered at $(b,l) = (20^{\circ},10^{\circ})$, via pair annihilation of 1.5~TeV DM particles to $e^+e^-$. The background is extracted from Ref.~\cite{Yuan:2017ysv}. }
\label{fig:sec-el}
\end{figure}

\begin{figure}[t]
\centering
\includegraphics[scale=.20]
{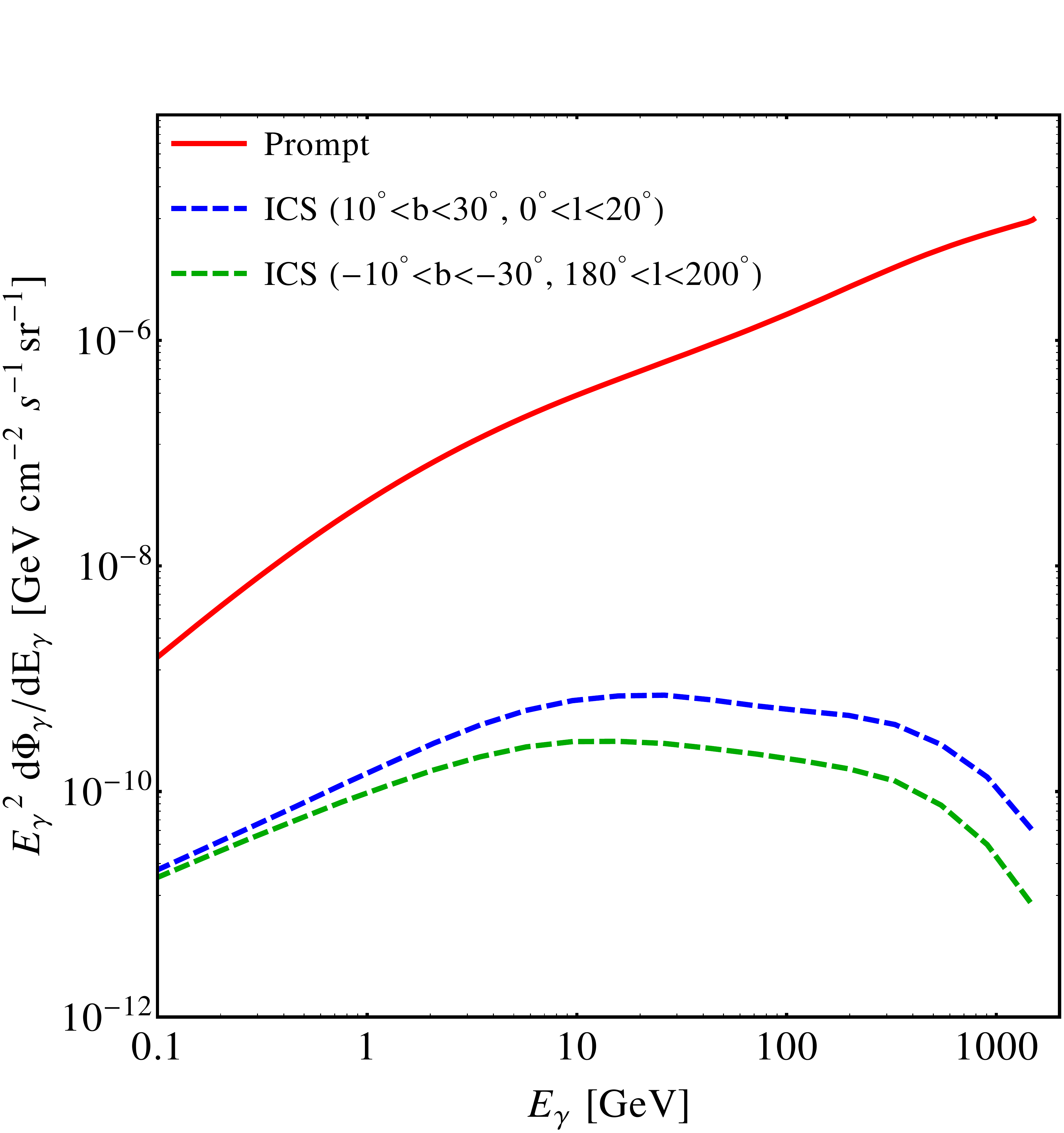}
\caption{A comparison of the prompt and ICS photon spectra arriving from the direction of  a clump at a distance of 0.1~kpc and centered at $(b,l) = (20^{\circ},10^{\circ})$, within a $20^{\circ} \times 20^{\circ}$ region.   For ICS we also show the flux in a region of the same size but in a direction diametrically opposite to that of the clump. The DM parameters are the same as Fig~\ref{fig:sec-el}. }
\label{fig:ICS}
\end{figure}

In Fig.~\ref{fig:sec-el} we show the secondary electron flux for the $e^+e^-$ annihilation channel with $m_X=1.5$~TeV and compare it with DAMPE data. We adopt an NFW profile~\cite{nfw} for the clump with the parameters of the profile calculated by fitting the best-fit mass ($5 \times 10^6 M_{\odot}$) and $\cal L$ ($3.5 \times 10^{64}$ GeV$^2$/cm$^{3}$) presented in Table III of Ref.~\cite{Yuan:2017ysv}. We incorporate the background by digitizing the left panel of Fig.~7 of Ref.~\cite{Yuan:2017ysv}. 
In Fig.~\ref{fig:ICS}, we plot the prompt photon spectrum and the ICS photon spectrum arriving from the direction of the clump for the same best fit point.
We see that the ICS photon flux is negligible compared to the prompt flux.  This is largely because at these energies ICS is dominated by interactions with IR light and starlight, which are not concentrated in the region of our search~\cite{Vincent:2010kv}.

\begin{figure}[t]
\centering
\includegraphics[scale=.20]{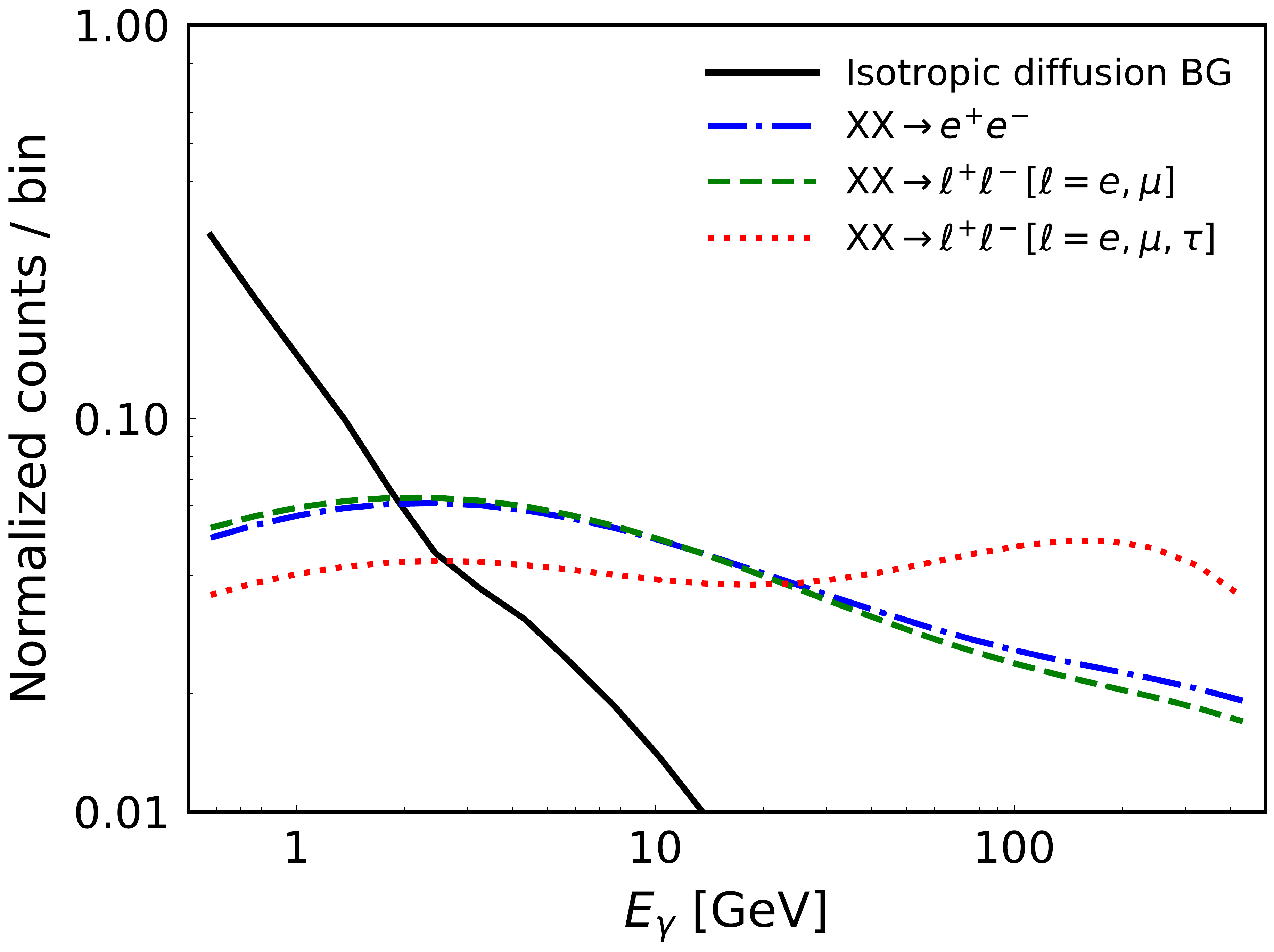}

\caption{A comparison of the expected photon counts per energy bin for the isotropic background, and for pair annihilation of DM particles of mass 1.5~TeV annihilating to various leptonic channels. All the spectra are normalized to unity. }
\label{fig:spec}
\end{figure}

In Fig.~\ref{fig:spec} we plot the photon spectrum for the isotropic background using the default parameters, as well as the gamma-ray spectra from DM annihilation for three leptonic final states for $m_X=1.5$ TeV. All the spectra in Fig.~\ref{fig:spec} are normalized to unity. The significantly different signal shapes for DM annihilation versus the expected isotropic background indicate that allowing the normalization of the isotropic template for each ROI to vary will not hide a signal of DM annihilation.  This analysis was repeated for several ROIs, keeping the normalization of the isotropic template fixed to its {\tt iso$\_$P8R2$\_$ULTRACLEANVETO$\_$V6$\_$v06.txt} prescribed value. 

Note that we do not attempt to simultaneously fit the DM annihilation signal along with the background; to do so, one must assume a spatial model
for the DM clump.  Instead, we fit a background
model to the data using both energy and angular information, extract the residual of the fit to the energy spectrum, and fit that spectral residual to a DM scenario.

\section{ Fitting DM annihilation to the residual flux in each ROI}
 We outline below our spectral analysis of the residual photon spectrum from each ROI.
For each DM scenario, we determine the photon spectrum $dN / dE_\gamma$ as described above.
We treat
the overall normalization, $\mathcal{N}$, of the photon flux as a free parameter, which we fit in our analysis.
We evaluate the number of photon counts per energy bin by integrating the photon spectrum over the width of each of the 24 bins obtained from the {\tt Fermipy} output. The overall normalization $\mathcal{N}$ is a product of the thermally averaged total DM annihilation cross section ($\braket{\sigma v}$), the unaveraged astrophysical $J$-factor of the clump, a prefactor $(8 \pi m_X^2)^{-1}$, and the effective area and the exposure time of the LAT detector. 

Finally, we perform a global fit of the spectral residual without and with a DM component and perform a $\chi^2$ analysis to evaluate the evidence for a DM clump of angular size $\mathcal{O}(10^\circ)$, with
\begin{equation}
\resizebox{.48 \textwidth}{!} 
{
$\chi^2 = 2 \sum_{i=1}^{24} \bigg[ N_{{BG}_i}+N_{{DM}_i} - N_{{Obs}_i} + N_{{Obs}_i} \ln{ \Big(\frac{N_{{Obs}_i}}{N_{{BG}_i}+N_{{DM}_i}} \Big)} \bigg]$\,,
}
\label{chi2}
\end{equation}
where $N_{{Obs}_i}$, $N_{{BG}_i}$  and $N_{{DM}_i}$ are the number of observed photons, the expected number of photons from astrophysical backgrounds, and the expected number of photons from DM annihilation, respectively, in the $i^{\rm{th}}$ bin; the last term in Eq.~(\ref{chi2}) vanishes if $N_{{Obs}_i}=0$.
For each ROI and each annihilation channel, we fit $\mathcal{N}$ in the range  
$\mathcal{N} = [10^{-20},10^{20}]$ with the DM mass fixed at 1.5~TeV, which is the best-fit value required to explain the DAMPE excess; we have checked that varying $m_X$ 
in the range $[1200, 1700]$~GeV improves the fit only marginally.



\begin{figure*}[t]
\centering
\includegraphics[scale=.17]{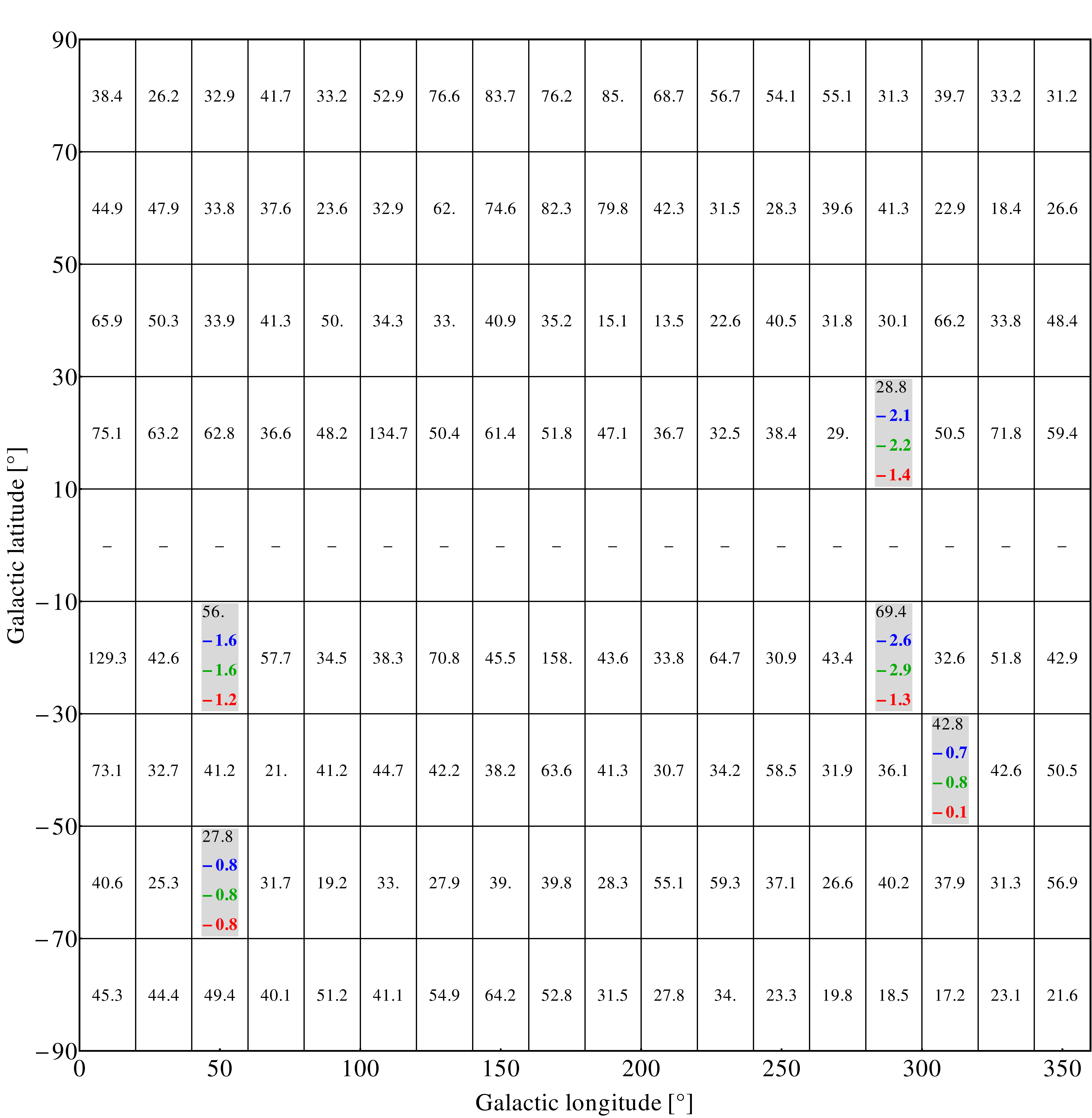}

\caption{The $\chi^2$ values for each ROI without dark matter annihilation. ROIs that show a preference for a DM component are shaded and include the $\Delta \chi^2=\chi^2_{DM}-\chi^2_{no\, DM}$ values for {\it Case I} (blue), {\it Case II} (green), and {\it Case III} (red), from top to down. The DM fits have 23 degrees of freedom (dof) with $m_X=1.5$~TeV. A 2$\sigma$~C.L. improvement is defined by $\Delta \chi^2=-4$ for one free parameter; this criterion does not apply for ROIs with a $\chi^2$/dof that is large compared to unity. Note that the latitudes and longitudes correspond to the center of the ROI, and do not represent the the span of latitudes and
longitudes of the ROI.}
\label{fig:red-chi2-ee}
\end{figure*}
\section{Results}
In Fig.~\ref{fig:red-chi2-ee} we provide the $\chi^2$ values for each ROI without a DM component. The inclusion of a DM contribution does not change the $\chi^2$ values significantly for any ROI. However, there are five~ROIs for which the inclusion of a DM component improves the $\chi^2$ by at most 2.9. ROIs that show a preference for a DM component are shaded 
and include the $\Delta \chi^2=\chi^2_{DM}-\chi^2_{no\, DM}$ values for {\it Case I} (blue), {\it Case II} (green), and {\it Case III} (red).
Three of these five ROIs are centered at $b = \pm 20^\circ$. 

Rather than floating the normalization of the isotropic background, one could instead fix the normalization to its default value; in some cases, the isotropic normalization found by
the fit in a particular ROI exceeds the default value,  and therefore may be concealing a DM signal.  We have checked this possibility for all five ROIs that show a mild preference for DM,  and for four that do not, and find a qualitatively similar result to that presented in Fig.~\ref{fig:red-chi2-ee}; no ROI is found that significantly prefers the inclusion of a DM component.


The ROI centered at $(b,\, l) = (-20^\circ, \, 290^\circ)$ shows the biggest improvement on inclusion of DM annihilation. The $\chi^2$ for this ROI without a DM contribution is 69.4. If a photon flux from DM annihilation is added to the background, the $\chi^2$ improves to 66.8, 66.5, and 68.1, with $\mathcal{N} = \, $ 80, 90, and 20, for {\it Cases-I}, {\it II}, and {\it III}, respectively. 

\section{Conclusion}
If the peak seen in the DAMPE electron spectrum is a result of the annihilation of
$\sim 1.5$~TeV DM particles to charged lepton pairs within a nearby dark matter clump, then the co-occurring gamma-ray emission arising from
bremsstrahlung that points back to the clump may be detectable via its distinct spectral shape.  We have shown that for the properties
of the clump needed to explain the DAMPE peak, the associated photon excess
is not discernible in \fermi~data. 

An interpretation of our null result as disfavoring a dark matter explanation should be accompanied by two caveats.
If DM annihilation occurs in a local overdensity which
includes the solar system, then
a photon excess would be absorbed in the isotropic background, leaving no residual.
A less serious proviso is that we searched for a clump by dividing the
sky into regions with an angular width of $20^\circ$ around the ROI center. If the clump did not lie entirely within a region, statistical
evidence for it would be diluted, and instead better evidence for dark matter
would be obtained by centering the ROI differently.
Indeed, the two neighboring ROIs show some improvement in the fit when a DM
component is included. However, the improvements in the fits are sufficiently 
small that we do not expect that moving the ROI will increase the evidence for a dark matter
clump.

Our main result is that we have
found no ROIs worthy of a closer look.

\vspace{0.2 in}
{\bf Acknowledgments.}
We thank K.~Boddy for participating in the early stages of this work.
We thank A~Drlica-Wagner, D.~Hooper and T.~Linden for useful discussions.
The work of T.G.~and J.K.~is supported in part by NSF CAREER Grant No.~PHY-1250573.
The work of D.M.~is supported in part by DOE Grant No.~DE-SC0010504.
The work of P.S.~is supported by NSF Grant No.~PHY-1720282.
We used the Texas A$\&$M University Brazos HPC, and University of Utah CHPC cluster resources to analyze \fermi~data.

\vspace{0.2 in}

\section*{Appendix}

We present the detailed pipeline of our background modeling with {\tt Fermipy}.
The {\tt Fermipy} methods and configuration parameters are italicized. The first step of our analysis for each ROI is to select a background model including {\tt 3FGL} point sources within a width of  $30^\circ $ around each ROI center, the interstellar emission model, and the corresponding isotropic component.
We perform a global fit of the normalizations of all model components. We preserve the power-law, log-parabola, or power-law with exponential cut-off nature of the spectral models of point sources, along with their indices, as described in the {\tt 3FGL} catalog.

Next, we use the {\it find$\_$sources} method to add new sources, with test statistic
$TS > 25$, to the background model, assuming a point source model for test sources with a power-law spectrum of index 2. This process scans the TS map of the ROI under investigation to find peaks in the TS map with the above condition, and adds a new point source at that location. It runs iteratively, and after each iteration, it generates a new TS map with the refined background model including the point sources identified in the previous iteration. The algorithm continues until no new peaks are observed with $TS > 25$.

After adding new sources to our model, we free the normalizations, as well as spectral indices where appropriate, of all background components within a $10^\circ$ radius of each ROI center and repeat the global fit of all free parameters. This concludes our background modeling for each ROI.

\end{document}